\documentclass[reprint,amsmath,amssymb,aps,twocolumn]{revtex4-2}

\usepackage{import}
\usepackage{amsmath, amssymb}
\usepackage{dsfont}
\usepackage[colorlinks=true,linkcolor=blue,citecolor=cyan]{hyperref}
\usepackage{graphicx}
\usepackage[dvipsnames]{xcolor}
\usepackage{natbib}
\usepackage{bookmark}
\usepackage{eqnarray}
\usepackage{placeins}
\usepackage{soul}
\usepackage{siunitx}
\usepackage{subcaption}
\usepackage{float}
\usepackage{multirow}
\usepackage{bm}

\usepackage{array}
\usepackage[utf8]{inputenc}

\begin{document}

% \pagenumbering{gobble}

\title{Turbulent relaxation patterns in SOL plasma}

\author{
\mbox{R. Varennes$^1$}, 
\mbox{G. Dif-Pradalier$^2$}
\mbox{P. Ghendrih$^2$},
\mbox{V. Grandgirard$^2$},
\mbox{O. Panico$^3$},
\mbox{Y. Sarazin$^{2}$},
\mbox{E. Serre$^4$},
\mbox{D. Zarzoso$^4$}
}
\affiliation{%
% \begin{singlespace}
 $^1$School of Physical and Mathematical Sciences, Nanyang Technological University, 637371, Singapore. \\
 $^2$CEA, IRFM, F-13108 Saint-Paul-Lez-Durance, France. \\
 $^3$Laboratoire de Physique des Plasmas, CNRS, IP Paris, Universite Paris-Saclay, 91128 Cedex, Palaiseau, France. \\
 $^4$ Aix-Marseille Université, CNRS, Centrale Marseille, M2P2, UMR 7340 Marseille, France.
% \end{singlespace}
}%

\date{\today}

\begin{abstract}
Relaxations of localized over-density in a plane transverse to the magnetic field are numerically investigated under the effect of drift-wave and interchange drives in SOL conditions.
Such a controlled departure from thermodynamic equilibrium allows the investigation of fundamental processes at play in cross-field transport.
Interchange instabilities generate ballistic outward radial flux with low amplitude zonal flow patterns, whereas drift-wave instabilities result in symmetric radial flux with large amplitude zonal flow patterns. 
When both instabilities are considered, the combined effects tend to favor drift-waves, leading to a weaker outward flux with larger zonal flow patterns.
\end{abstract}

\maketitle

% !TeX spellcheck = en_GB

%!!!!!!!!!!!!!!!!!!!!!!!!!!!!!!!! SECTION !!!!!!!!!!!!!!!!!!!!!!!!!!!!!!!!!!!!!!!!!!!!!!!!!!!!!!!!!!!!!!!!!!!!!!!!!!!!!!!!!!!!!!!!!!
\section{Introduction}
\label{section:introduction}
Understanding SOL turbulent transport as resulting from relaxation events has been a major breakthrough \cite{Sarazin1997, Sarazin2000, Krasheninnikov2001}. 
This has opened the way to a vast literature and comparison to experimental evidence.
The body of papers being too long to review here, the reader is referred to the reviews by \textit{D'Ippolito} et al. \cite{DIppolito2011} and \textit{Krasheninnikov} et al. \cite{Krasheninnikov2008}.
One of the groundbreaking discoveries has been the change of turbulence forcing from that governed by fixed gradients, i.e. gradient-driven, to that governed by prescribed average fluxes, i.e. flux-driven. 
Indeed, in the standard description of near thermodynamic equilibrium transport, the fluxes are written in terms of a linear dependence on the gradients, with constant matrix elements proportional to various diffusivities. 
The departure from thermodynamic equilibrium, characterized by the various gradients, induces fluxes that tend to restore thermodynamic equilibrium. 
With weak fluctuations of both gradients and fluxes, the linear relationship holds for the mean fields and is readily interpreted as gradients driving fluxes. 
However, in most systems the actual drive out of thermodynamic equilibrium are fluxes and particular experimental skills are required to fix the gradients. 
In most situations, the gradients exhibit a dynamic response to the fluxes that drive the system out of equilibrium. 
In most cases, and in particular in magnetic confinement devices dedicated to nuclear fusion experiments, sources are controlled and sustain fluxes that in turn generate gradients and bring the system out of equilibrium. 
Because competing loss mechanisms exist, the onset of turbulent transport only occurs above a threshold in the source magnitude. 
\newline
An important part of the associated transport is carried by intermittent relaxation events, which are called \textit{avalanches} in this paper, leading to dense and coherent filamentary structures - often referred to as \textit{blobs}.
This intermittent radial transport occurs ballistically at relatively constant velocity \cite{Choi2024}.
The turbulent transport paradigm then shifts from diffusive to ballistic. 
In such systems, constant sources only prescribe the mean fluxes such that large fluctuations of both gradients and fluxes are permitted. 
The dynamics bear similarities with Self-Organised Criticality \cite{Bak1987}. 
\newline
This intermittent transport carried by blobs is supported by many experimental observations \cite{Ayed2009, Bisai2022, Boedo2003, Maqueda2010, Muller2007, Nold2010, Terry2005, Zweben2021, Katz2008, Choi2024}.
These experimental evidences triggered many theoretical and numerical works \cite{Benkadda1994, Krasheninnikov2001, Decristoforo2021, Ghendrih2018, Ghendrih2022, Losada2023}.
In particular, multiple simplified sets of equations \cite{Ghendrih2022, Bian2003, Angus2012, Angus2012a, Easy2014} modeling the SOL region have been proposed and solved numerically, which successfully replicate intermittent transport triggered by avalanches and carried by dense blobs.  
One of the main ingredients to recover this behavior is the interchange instability \cite{Garbet1991, Nedospasov1993, Benkadda1994a}.
When accounting for the resistive parallel losses in the SOL, a finite phase shift between density and electric potential can emerge, which can trigger the drift-waves instability.
In a previous paper \cite{Ghendrih2022}, the dispersion relation of a simplified system of equations including both the interchange and drift-wave instabilities has been used to devise two numerical scenarios where each instability has been simulated separately with a similar growth rate.
These cases, along with another one where both instabilities contribute equally to the growth rate, were employed to gain new insights into statistically relevant properties of transport, e.g. the SOL width or the turbulent intensity.

In this paper, these scenarios are used to study a single relaxation event by letting a unique turbulent structure, i.e. a blob, evolve under the drift-wave and interchange instabilities.
Key properties, specific to each instability, of the resulting isolated turbulent structures are determined.
A particular attention is given to the role of each instabilities in the zonal flow generation and the outward particle flux.
When both instabilities are accounted for on the same footing, the combined effects are observed to be more closely aligned with the drift-wave behavior.
Despite extensive research on isolated relaxation events \cite{Kendl2018, Bian2003, Krasheninnikov2001, Angus2012, Angus2012a, Easy2014}, the decoupling of each instability in such a controlled manner has not been done previously.
% In addition, this work is also meant as a pioneering study for establishing a systematic numerical method for detecting structures in turbulent plasmas.
% Structure identification is appealing for filtering turbulent events in simulations and experimental data. 

The remainder of the paper is organized as follows.
The governing equations and the linear properties of the interchange and drift-waves instabilities are presented in Section \ref{section:SOL_turbulence_model}.
The simulation framework, including a description of the \textsc{tokam2d} code and the associated non-linear simulations considered, is detailed in Section \ref{section:numerical_setup}.
The response to prepared initial conditions with both instabilities is addressed in Section \ref{section:ballistic_density_propagation}. 
%The analysis of characteristic turbulent simulations are found in Section \ref{section3_transport_patterns_in_turbulent_SOL_plasma}. 
The discussion and conclusion in Section \ref{section:discussion_and_conclusion} closes the paper. 
% !TeX spellcheck = en_GB
%
\section{SOL turbulence model}
\label{section:SOL_turbulence_model}
% 

% In this section, the aim is to identify the key processes, and the characteristic structures they generate, which are combined to sustain self-organised turbulent transport. 

\subsection{Transverse transport model}
\label{subsection: Transverse transport model}
Plasma turbulence is investigated with a minimum transport model, arguing that the robust physics at play will be qualitatively addressed. 
These models couple at most two fields and use fluid equations restricting the phase space to the directions transverse to the magnetic field, typically $r$ a radial and $\theta$ an angle coordinate. 
While details on the present model used throughout this paper and its derivation can be found in \cite{Ghendrih2022}
%and Appendix \ref{section: The 2-field turbulence model}
, hasty readers can refer to the brief following overview. 
In this model, the linear drive for instability is either that of drift wave turbulence, as investigated in the seminal paper \cite{Hasegawa1983}, or interchange-like \cite{Garbet1991, Nedospasov1993, Sarazin1997}.
A modulational Kelvin-Helmholtz instability can also be triggered \cite{Norscini2015, Norscini2022}. 
Parallel transport is taken into account by loss terms that are simplified to only linearly depend on the chosen fields, as discussed in \cite{Ghendrih2022}, and is akin to the drift-waves model in \cite{Berk1991}.
This transport model takes the form of governing equations for the vorticity $W$ and the particle density $n$ - alternatively the scalar pressure $p = nT$ proportional to the internal energy density, or, combining these equations, a transport equation for the thermal energy $T$. 
Cold ion limit is assumed so that the plasma thermal energy $T$ is that of electrons, $T=T_e$. 
In this limit, the Boussinesq approximation allows one to determine the electric potential $\phi$, given the vorticity via $W = \boldsymbol{\nabla}_\perp^2 \phi$.  
The drift expansion is used to determine the transverse transport, at the lowest order governed by the electric and diamagnetic drift velocities, and at higher order by the ion polarisation velocity. 
In these assumptions, the transverse transport model for plasma turbulence is akin to the Rayleigh-B\'enard 2D turbulence model \cite{Normand1977} for neutral fluids, as underlined in \cite{Ghendrih2003, Wilczynski2019}. 
Transverse transport then appears as a competition between convection and diffusion. 
Above a threshold in the departure from thermodynamic equilibrium, a bifurcation takes place from purely diffusive transport to convection-dominated transport. 
%
%We consider here the linearised expressions depending on the density and electric potential.
Finally, normalizing the equations, simplifying them consistently with the drift ordering, and stepping to a slab geometry, with $x = (r-a )/\rho_0$ the radial coordinate, and $y = a\theta/\rho_0$ the poloidal coordinate, one obtains:
\begin{subequations}
	\label{system: TOKAM2D}
	\begin{align}
		\partial_t n + \frac{1}{L_{\rm n}} \partial_y\phi + \Big[\phi,n\Big] - D\nabla_\perp^2n &= S_{\rm n} -\sigma_{\rm n,n} n + \sigma_{\rm n,\phi} ~\phi
		\label{equation: evolution n}\\
		\partial_t W + g \partial_yn + \Big[\phi,W\Big] -\nu\nabla_\perp^2W &= S_W - \sigma_{\rm \phi,n} n + \sigma_{\rm \phi,\phi} \phi
		\label{equation: evolution W}
	\end{align}
\end{subequations}
The density is normalized by a characteristic density, while the electric potential is normalized by $T_e /e$. 
As indicated by the definitions of $x$ and $y$, length scales are normalized by the reference Larmor radius $\rho_0$. 
Time is normalized by a reference ion cyclotron period $\Omega_0^{-1}$.  
The density and vorticity diffusion coefficients $D$ and $\nu$ are assumed stationary and homogeneous.
The RHS of both equations are analogous and contain a density/vorticity source as well as terms proportional to $\sigma_{i,j}$ which account for the parallel losses.
Finally, the Poisson brackets are defined by $[f,g]  = \partial_x f\partial_y g - \partial_yf \partial_x g$.

The simulation domain is restricted to a region with a reduced poloidal extent on the low field side midplane, which yields the interchange term proportional to $g$ in the vorticity equation. 
With this simplification, the poloidal direction is homogeneous.
The system Eqs(\ref{system: TOKAM2D}) allows addressing two means of driving the density field out of equilibrium and sustaining instabilities that drive SOL turbulence: either the particle source terms $S_{\rm n}$ in the flux driven regime, or the linear term $(1/L_{\rm n}) \partial_y\phi$, which is proportional to the length scale of the density gradient defined by $L_{\rm n} = - (\nabla_x \ln{n_0})^{-1}$, in the gradient driven regime. 
In the former case, the mean density gradient is a result of the turbulent transport while, in the latter case, one focuses on density fluctuations $n$ such that the total density reads $n_{\rm tot}(x,y,t) = n_0(x) + n(x,y,t)$ and the mean turbulent particle flux is the output. 
In the edge and SOL plasma, where the mean fields exhibit poloidal and radial variations, the gradient-driven approach is not relevant. 
One can also drive the electric potential out of equilibrium by assuming a source $S_{\rm W}$ generating a radial stratification or a cubic expansion of the flux-surface averaged electric potential $\langle \phi \rangle_y(x)$ yielding a poloidal Doppler velocity $\nabla_x\langle \phi \rangle_y(x)$, possibly a radial dependence of the vorticity $\langle W \rangle_y = \nabla_x^2\langle \phi \rangle_y(x)$, and a gradient of this vorticity $\nabla_x\langle W \rangle_y$ (see 
%Appendix \ref{section: The 2-field turbulence model} and 
\cite{Ghendrih2003} for the stability analysis with this particular expansion). \\
%\textcolor{red}{(On peut aussi driver le potentiel avec seulement le terme d'interchange me semble-t-il.)}
%
In the next section, the dispersion relation of a simplified system Eqs(\ref{system: TOKAM2D}) is derived to extract key properties specific to the drift-wave and interchange instabilities.
\subsection{Dispersion relation}
\label{subsection:dispersion_relation}
%
%However, we shall consider it to perform the linear analysis.  
From the governing equations in the system Eqs(\ref{system: TOKAM2D}), the dispersion relation is derived by considering a steady-state solution such that $\langle \phi \rangle_y = 0$, typically for $S_W=0$ - therefore excluding the modulational Kelvin-Helmholtz instability - and with a gradient-driven approach for the density - hence for $S_{\rm n}=0$ and finite $(1 / L_{\rm n})$.
In a linear framework, hence simplifying the Poisson brackets in Eqs(\ref{system: TOKAM2D}), the coupling between the two equations is governed by the terms $(1/L_{\rm n}) \partial_y\phi$ and $\sigma_{\rm n,\phi} \phi$ in the density evolution equation Eq(\ref{equation: evolution n}), and the two terms $g\partial_y n$ and $ \sigma_{\rm \phi,n} n$ in the vorticity equation Eq(\ref{equation: evolution W}). 
Without these coupling terms, the remaining terms enforce an exponential decay, as readily expected for damping processes. The coupling terms are therefore mandatory to drive turbulent transport.
The growth rate $\gamma$ is computed with the linearised equations in Fourier space.
The Fourier transform of the vorticity $\widehat{W}$ then verifies $\widehat{W} = -k^2\widehat{\phi}$ where $k^2 = k_x^2 + k_y^2$, $k_x$ and $k_y$ being the wave vectors in the $x$ and $y$ directions respectively. 
The growth rate $\gamma$ is defined as $\partial_t \widehat{n}/\widehat{n} = \partial_t\widehat{\phi}/\widehat{\phi}$, therefore:
\begin{subequations}
	\label{system: linear TOKAM2D Fourier}
	\begin{align}
		\gamma ~\widehat{n} + \frac{ik_y}{L_{\rm n}}\widehat{\phi}  + D k^2\widehat{n} &= -\sigma_{\rm n,n} \widehat{n} + \sigma_{\rm n,\phi} ~\widehat{\phi}
		\label{equation: evolution n Fourier}\\
		\gamma ~\widehat{\phi} - i\frac{k_y}{k^2} g \widehat{n}  + \nu k^2\widehat{\phi} &= -\frac{\sigma_{\rm \phi,\phi}}{k^2} \widehat{\phi} + \frac{\sigma_{\rm \phi,n}}{k^2} \widehat{n}
		\label{equation: evolution W Fourier}
	\end{align}
The dispersion relation determines the condition to achieve a solution different from the trivial solution $\widehat{n} =0$ and $\widehat{\phi}=0$. 
For such modes, the growth rate is determined by the following second-order equation:
	\begin{align}
		\big(\gamma +A_{\rm n}\big)\big(\gamma +A_{\rm \phi}\big) -B_{\rm n} B_{\rm \phi} = 0
		\label{equation: dispersion relation}
	\end{align}	
\end{subequations}
where:
\begin{subequations}
	\begin{align}
		A_{\rm n}= D k^2 +  \sigma_{\rm n,n}
		\hspace{3em}&; \hspace{3em}
		A_{\rm \phi}= \nu k^2 +  \frac{\sigma_{\rm \phi,\phi}}{k^2}
		\label{equation: An et Aphi}\\
		B_{\rm n} = \frac{ik_y}{L_{\rm n}} - \sigma_{\rm n,\phi} 
		\hspace{3em}&; \hspace{3em}
		B_{\rm \phi} = -ig\frac{k_y}{k^2} - \frac{\sigma_{\rm \phi,n}}{k^2}
		\label{equation: Bphi Bn}
	\end{align}
\end{subequations}
The condition for instability $\text{Re}(\gamma)>0$ crucially depends on the product $B_{\rm n}B_{\rm \phi}$. 
For $B_{\rm n}~B_{\rm \phi}=0$, the solutions $\gamma = - A_{\rm n}$ and $\gamma = - A_{\rm \phi}$ are real and negative: there is no instability, the damping processes characterized by $A_{\rm n}$ and $A_{\rm \phi}$ govern stability. 
%The complete instability analysis is recalled in Appendix \ref{subsection: Dispersion relation Appendix}. 
The instability condition is found to be:
\begin{subequations}
	\begin{align}
	\text{Re}\big(B_{\rm n} B_{\rm \phi}\big) &+ \frac{\text{Im}\big(B_{\rm n} B_{\rm \phi}\big)^2}{\big(A_{\rm n} + A_{\rm \phi}\big)^2} > A_{\rm n} A_{\rm \phi}
	\label{equation: instability constraint}\\
	\text{Re}\big(B_{\rm n} B_{\rm \phi}\big) &= \frac{\sigma_{\rm \phi,n}\sigma_{\rm n,\phi}}{k^2} + \frac{k_y^2}{k^2}\frac{1}{L_{\rm n}}g
	\label{eq:ReB_nB_phi}\\
	\text{Im}\big(B_{\rm n} B_{\rm \phi}\big) &=\frac{k_y}{k^2} \left(g \sigma_{\rm n,\phi} - \frac{1}{L_{\rm n}}\sigma_{\rm \phi,n}\right)
	\label{eq:ImB_nB_phi}
	\end{align}
\end{subequations}
When setting $g=0$ and $\sigma_{\rm \phi,n}=0$, the instability condition Eq(\ref{equation: instability constraint}) cannot be met. 
From the linear point of view, one thus identifies:
\begin{itemize}
    \item the interchange instability \cite{Garbet1991,Nedospasov1993} with $g>0$ and $\sigma_{\rm \phi,n}=0$ on the one hand;
    \item the drift wave instability \cite{Hasegawa1978, Hasegawa1983} with $g=0$ and $\sigma_{\rm \phi,n}>0$ on the other hand.
\end{itemize} 
Further simplifying these two cases by setting $\sigma_{\rm n,\phi} =0$ greatly facilitates the analysis and yields the following key properties of the instability thresholds:
\begin{subequations}
	\begin{align}
		\text{Re}\big(B_{\rm n} B_{\rm \phi}\big) &=  \frac{k_y^2}{k^2} \frac{1}{L_{\rm n}} g
		\label{eq:ReB_nB_phi_simplified}\\
		\text{Im}\big(B_{\rm n} B_{\rm \phi}\big) &= - \frac{k_y}{k^2} \frac{1}{L_{\rm n}} \sigma_{\rm \phi,n}
		\label{eq:ImB_nB_phi_simplified}	
	\end{align}
\end{subequations}
For the SOL interchange case, i.e. $g>0$ and $\sigma_{\rm \phi,n}=0$, the product $B_{\rm n}B_{\rm \phi}$ is real and found positive for $(1/L_{\rm n}) = - \nabla_x \ln{n_0} > 0$, i.e. negative background gradient, which is therefore a necessary condition for the instability. 
The buoyancy effect governed by $g$ can only drive the SOL interchange instability on the low field side \cite{Garbet1991,Nedospasov1993}. 
Turbulent transport is then ballooned to the low field side similar to that evidenced experimentally \cite{LaBombard2004, Asakura2007, Gunn2007}.
In the case of the drift wave instability, i.e. $\sigma_{\rm \phi,n}>0$ and $g=0$, the product $B_{\rm n}~B_{\rm \phi}$ is purely imaginary. 
The threshold constraint is then a function of $(1/L_{\rm n})^2$ with no constraint on the sign of the density gradient. 
With this symmetry, the turbulent transport is identical in the low and high field sides of the device. 
In both cases, convective $E\times B$ transport is triggered past a threshold on the density gradient ensuring that the drive is larger than the damping processes.
In other words, the source must overcome the transport capability of the diffusive and parallel transport channels to trigger the convective instability. 
For the interchange case, the growth rate is real so that the density and radial projection of the electric drift velocity $v_{Ex} = - \partial_y \phi$ are in phase. 
This phasing maximizes the particle flux. 
Conversely, for the drift-waves case, the growth rate exhibits an imaginary part, which stands for the mode frequency, so that the density and $v_{Ex}$ fluctuations are no longer in phase. 
This reduces the transport efficiency of the electric drift turbulent motion. \\
 
One can also note that $\text{Re}\big(B_{\rm n} B_{\rm \phi}\big)$ and 	$\text{Im}\big(B_{\rm n} B_{\rm \phi}\big)$ depend on either $k_y^2/k^2$ or $k_y^2/k^4$. 
An important effect is that the drive for $k_y=0$ modes is null, and since the damping terms are not vanishing, all modes $k_y=0$ are stable. 
To complete this case, one can remark that for $\sigma_{\rm n,\phi}>0$ and $k_y=0$ the instability constraint Eq(\ref{equation: instability constraint}) takes the form $\sigma_{\rm \phi,n}\sigma_{\rm n,\phi} > k^2 A_{\rm n}A_{\rm \phi}$. 
One readily finds that $k^2 A_{\rm n}A_{\rm \phi} = \sigma_{\rm \phi,n}\sigma_{\rm n,\phi} + \sigma_{\rm \phi,n} D k^2 + \sigma_{\rm n,\phi}\nu k^4 + D\nu k^6$ so that the instability constraint cannot be satisfied for the $k_y$ modes. 
A general result is therefore that all modes $k_y=0$ are stable. 
Related to this point, one finds that the most unstable mode structure governed by the dependence on $k_y /k$ enforces $|k_y| = |k_x|$ to maximize the instability driving term.\\

In the instability constraint Eq(\ref{equation: instability constraint}), the damping terms are either governed by $A_{\rm n}A_{\rm \phi}$ in the interchange case or by $A_{\rm n}A_{\rm \phi}(A_{\rm n} + A_{\rm \phi})^2$ for the drift waves case. 
Since $A_{\rm \phi}$ exhibits a minimum for $k^4 = \sigma_{\rm \phi,\phi}/\nu$, one expects the damping process of both instabilities to exhibit a minimum, thus defining the most unstable wave vector $k_\star$. 
For standard simulation parameters, one finds $k_\theta \rho_0\approx 0.11$ for the interchange and $k_\theta \rho_0\approx 0.16$ for the drift waves, hence structures sizes of approximately $9 \rho_0$ and $ 6 \rho_0$ respectively. 
The minimum of $A_{\rm n} + A_{\rm \phi}$, which can be computed analytically, yields values comparable to that determined for the drift waves case. 

Finally, an important difference between interchange and drift waves with $\sigma = \sigma_{\rm n,n} = \sigma_{\rm n,\phi}= \sigma_{\rm \phi,\phi}= \sigma_{\rm \phi,n}$ is that, as $g$ is increased, $\gamma \to |k_y/k| ~\sqrt{g (1/L_{\rm n})}$. 
Consequently, the growth rate increases with $g$ provided $1/L_{\rm n} = -\nabla_x \ln{n_0} >0$. 
Conversely, as $\sigma$ increases, the electric potential tends to align on the density $n\to \phi$. 
In other words, the system tends to the well-known adiabatic limit with a null electron current. 
Within that limit, the growth rate $\gamma$ is negative. 
This stabilizing effect governed by the parallel current becomes more important than the destabilizing effect. 
The drift waves are first destabilized as $\sigma$ is increased and then stabilized when $\sigma$ is further increased.\\

In this section, some unique properties related to the interchange and drift-wave instabilities have been obtained from the dispersion relation obtained from the linearised governing equations.
These linear properties can be summarized as follows:
\begin{itemize}
    \item For the interchange instability: a positive growth rate is associated with a negative density gradient (i.e., $1/L_{\rm n}>0$), which acts as a necessary condition for instability. 
    This type of instability is predominantly influenced by the buoyancy effect, represented by the parameter $g$. 
    It drives turbulence on the low field side of the device, manifesting a ballooning of turbulent transport to these regions. 
    This instability primarily results in convection-dominated transport.
    The growth rate associated with this instability is real, leading to an enhanced particle flux due to the phase alignment between density fluctuations and the radial projection of the electric drift velocity.
    \item For the drift-wave instability: a positive growth rate does not require a specific sign of the density gradient, allowing for a more symmetric turbulence distribution across the domain. 
    This instability is characterized by a finite imaginary part of the growth rate, indicating an oscillatory behavior that can reduce transport due to the phase misalignment between density and electric drift velocity fluctuations. 
    The instability's threshold and growth rate are intricately tied to the values of $\sigma_{\rm \phi,n}$, which modulate the electric potential’s influence on the density field, and thus influence the overall stability.
\end{itemize}

These properties are of prime importance to interpret the non-linear simulations of these instabilities.
In the next section, the \textsc{tokam2d} code is introduced and the numerical setup of relevant non-linear cases is detailed.

\section{The numerical setup}
\label{section:numerical_setup}
\subsection{The TOKAM2D code}
\label{subsection: Numerical scheme for simulations}
%
%The reference mesh size is the normalized Larmor radius $\rho_0$. 
TOKAM2D is a spectral code \cite{Ghendrih2018, Ghendrih2022} with a simulation domain periodic in all directions.
Consequently, the simulations are performed in Fourier space, making extensive use of the fast Fourier transforms - which are efficiently executed on GPUs - to compute the non-linear terms. 
If the periodicity along the poloidal direction $y$ is quite standard, the one in the radial direction $x$ requires careful consideration.
Indeed, in flux-driven systems, on which this paper focuses, the density profile exhibits a maximum at the source location and a minimum at a location that depends on the transport properties. 
This structure has two implications.
First, these extrema also correspond to regions with vanishing particle fluxes. 
The minimum density region then appears to be non-turbulent, generating a buffer region inhibiting the cross-talk between turbulent regions.
Consequently, if the buffer region is wide enough, the apparent drawback of using a periodic radial direction is minimized. 
A second consequence is that both negative and positive mean density gradient regions, hence low and high field like regions, are simulated and can then be compared. 
Together with this interesting feature, the evolution in Fourier space offers an accuracy way superior to finite difference schemes in computing space derivatives.
\newline
The reference mesh size is the normalized Larmor radius $\rho_0$.
The time stepping is a standard Runge Kutta scheme of order 4. 
The typical time step is $1 / \Omega_0$, the reference cyclotron period of the ions at electron temperature. 
A typical number of time steps to reach steady state turbulent transport is then in the range of $10^4$ to $10^5$ time steps. 
Simulation can be started from either an analytical initial state or using a chosen state of the density and electric potential generated by another simulation. 
The latter procedure is of interest to reduce the transient to statistical steady state, in particular when scanning a control parameter. 
In the former case, a typical initial condition to investigate SOL turbulence is a constant density in both the $x$ and $y$ directions with an initial electric potential determined by a set of chosen small amplitude Fourier modes with random phases. 
In this paper, the initial conditions are chosen to generate an isolated avalanche.
\newline
When computing the evolution of system Eqs(\ref{system: TOKAM2D}) in Fourier space, the inversion of the equation $W =\nabla_\perp^2\phi$ leads to a factor $1/k^2$ on the right hand side of the equation determining $\widehat{\phi}$ as noticeable in the linearised form Eq(\ref{equation: evolution W Fourier}). 
A specific treatment is then made to determine $\widehat{\phi}(k_y=0,k_x=0)$. 
This is particularly important since the linearised parallel transport terms - those depending on the $\sigma$ conductivities - implicitly assume that the reference constant electric potential is null. 
For consistency, the code enforces that $\widehat{\phi}(k_y=0,k_x=0)$ remains null. 
\newline

\subsection{Initialization of simulations}

%Before investigating SOL turbulent transport we first address 
A simplified situation is considered to determine some basic properties of relaxation processed under the interchange and drift-waves instabilities. 
As stated in the introduction, an area of interest for these single relaxation events is to characterize their dynamics and properties in terms of transport efficiency. 
One can then determine means to identify them, and their weight in the overall transport, in more complex situations. 
% Such isolated relaxation events \cite{Kendl2018, Bian2003, Krasheninnikov2001, Angus2012, Angus2012a, Easy2014} can then be compared to actual turbulent transport regimes \cite{Sarazin1997, Ghendrih2003}. 
% In this simplified case, the background density is homogeneous and set equal to 1. 
%

\begin{figure}
	\centering
    \centerline{
        \includegraphics[width= 0.95\linewidth]{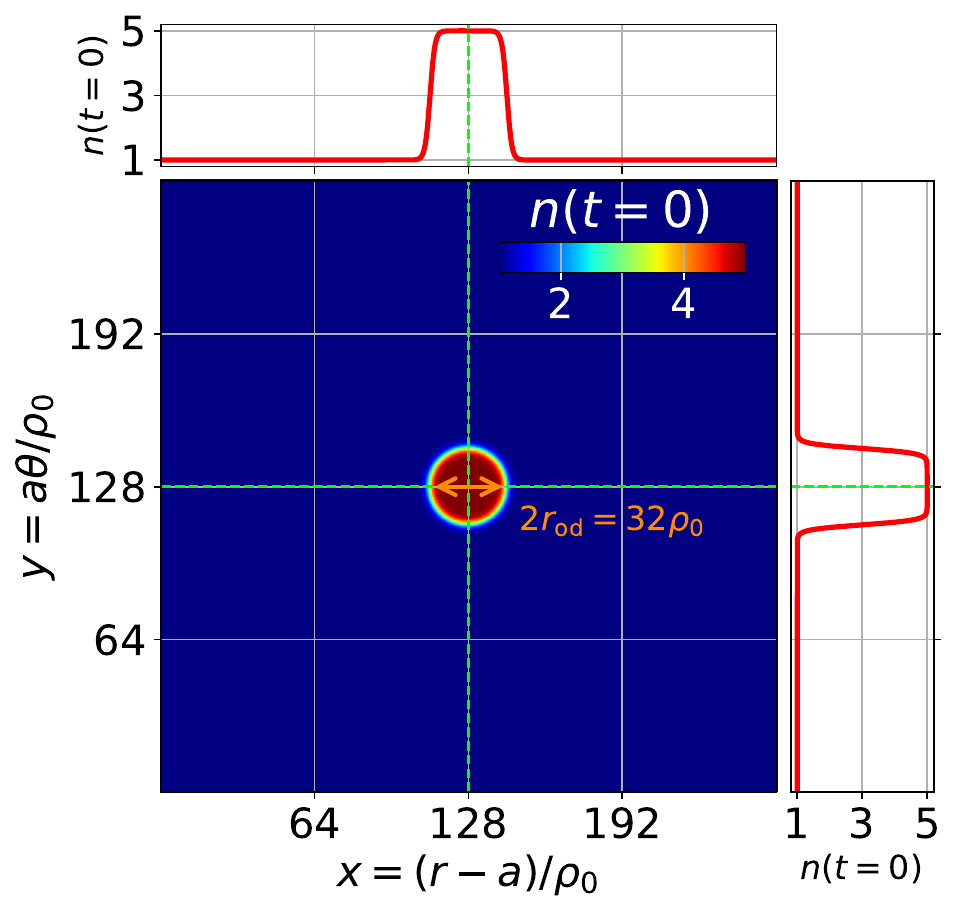}
    }
	\caption{\label{fig:initial_overdensity} \small Initial overdense region $n(x,y,t=0)$ initialized in the entire simulation domain for each case considered in this work. It is defined as $n(x,y,t=0) = n_{\rm bg} + \Delta n_{\rm od}(1 - \tanh((r-r_{\rm od})/L_{\rm od})/2$ where $r^2 = (x-x_{\rm od})^2 + (y-y_{\rm od})^2$ with the parameters defined in Table \ref{table:simulation_initial_conditions_blob}.}
\end{figure}

The initial conditions are set to create a circular overdense region characterized by its radius $r_{\rm od}$ and the variation scale $L_{\rm od}$ determining the transition between the overdense flat-top with height $~\Delta n_{\rm od}$ and the density background $n_{\rm bg}$, see Table \ref{table:simulation_initial_conditions_blob}.
More precisely, as shown in Fig.\ref{fig:initial_overdensity}, a hyperbolic tangent shape originating from the center $(x_{\rm od}, y_{\rm od})$ of the over-dense region is used such that the total initial density reads
\begin{equation}
    n(x,y,t=0) = n_{\rm bg} + \Delta n_{\rm od}(1 - \tanh((r-r_{\rm od})/L_{\rm od})/2
\end{equation}
where $r^2 = (x-x_{\rm od})^2 + (y-y_{\rm od})^2$. % with the parameters defined in Table \ref{table:simulation_initial_conditions_blob}. 
With a simulation spatial domain spanning $256 \rho_0$ radially and $256 \rho_0$ poloidally, the center of the initial overdense region is chosen at $x_{\rm od} = y_{\rm od} = 128 \rho_0$.
The density source is set such that $S_{\rm n} = \sigma_{\rm n,n}$ and is thus homogeneous in space and stationary.
The combination of this specific source $S_{\rm n}$ and the parallel density loss term $-\sigma_{\rm n,n} n$ ensures that the fixed point of the background density is 1. 

\begin{table}
	\renewcommand{\arraystretch}{1.5}
	\centering
	\begin{tabular}{|c|c|c|c|c|c|c|}
        \hline
		$~r_{\rm od} $ & $~L_{\rm od}$ & $~x_{\rm od}$& $~y_{\rm od}$ & $~S_{\rm n}$ & $~\Delta n_{\rm od}$  & $~n_{\rm bg}$  \\
        \hline
		$16 \rho_0$ & $~2 \rho_0$ & $128$ & $128$ &	$\sigma_{\rm n,n}$  &	$4 n_0$&	$1 n_0$  \\
        \hline
	\end{tabular}
\caption{\label{table:simulation_initial_conditions_blob} Initial conditions of the initial overdense region for isolated avalanche generation.}
\end{table}

\subsection{Simulation considered}

To be able to compare the relaxation of an isolated structure with each instability in a controlled way, three simulation cases are considered.

\begin{itemize}
    \item The ``interchange only" case (``I") where the only drive is the curvature term $g \partial_y n$ term. No phase shift between the density and the potential is considered such that the coupling coefficients $\sigma_{\rm n,\phi}=\sigma_{\rm \phi,n}=0$.
    \item The ``drift-waves only" case (``DW") where the only drive is the finite phase shift between potential and density, considered through $\sigma_{\rm n,\phi}=\sigma_{\rm \phi,n} \neq 0$. The curvature is set to zero, i.e. $g=0$.
    \item The ``combined interchange/drift-waves" case (``DW+I"), where $\sigma_{\rm n,\phi}=\sigma_{\rm \phi,n} \neq 0$ and $g \neq 0$.
\end{itemize}

The simulation parameters are chosen from a case selected in \cite{Ghendrih2022} designed such that the two instabilities are characterized by close to identical growth rates $\gamma \approx 4.3 ~10^{-4} ~ \Omega_0$. 
This growth rate is computed for $k_x \rho_0 = 5 \pi/256$, then selecting the maximum value when varying $k_y$. 

The result of the calculation of the growth rate $\gamma$ exhibits a small variation with $k_x$ for $1 \le k_x / k_{\rm min} \le 5$, leading to a typical size in the radial direction ranging from $16 \rho_0$ to $81 \rho_0$. 
In the poloidal direction, the structure size estimated with the linear analysis varies from $9~\rho_0$, for interchange, to less than $14~\rho_0$ for drift waves.
\newline
Regarding time scales, the characteristic time to reach steady state conditions is $ 1 / \sigma_{\rm n,n} \approx 20000 \Omega_0^{-1}$, which is one order of magnitude larger than the characteristic growth time $1/ \gamma\approx 2000 \Omega_0^{-1}$. 

\begin{figure*}
	\centering
    \centerline{
        \includegraphics[width=1.0\linewidth]{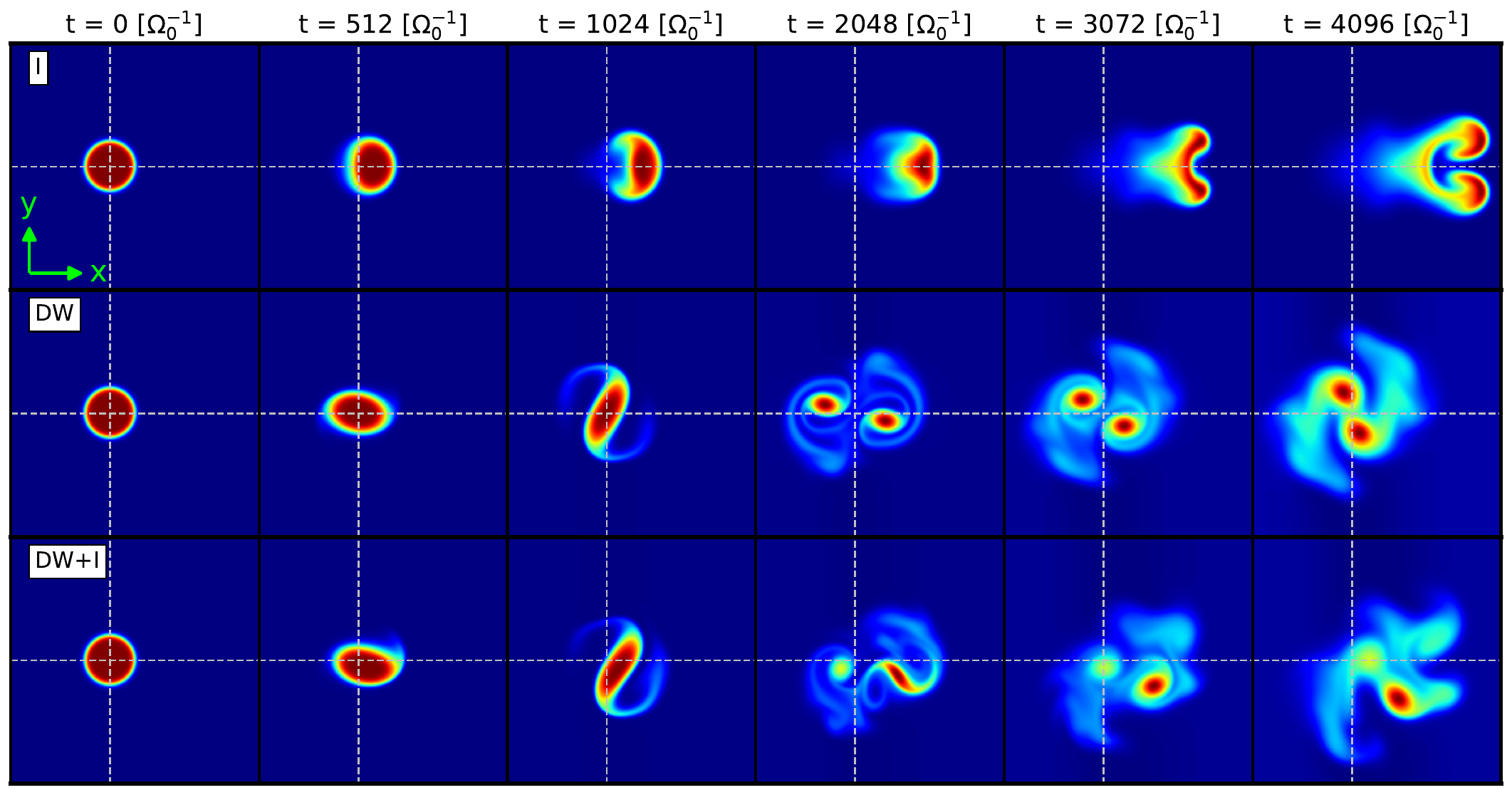}
    }
	\caption{\label{fig:structure_evolution_overview} \small Overview of the overdensity's structure evolution obtained with \textsc{tokam2d} with the parameters defined in Table \ref{table:drift_wave_interchange}.
    ``I" stands for interchange only, ``DW" for drift-waves only and ``DW+I" is a combination of these two cases.
    The physics underlying this evolution is discussed in Sec.\ref{section:ballistic_density_propagation}.}
\end{figure*}

A summary of the control parameters, simulation mesh, and details on the growth rate for each case is available in table \ref{table:drift_wave_interchange}. 
A visual overview showing the evolution of the density structures associated with each case is given in Fig.\ref{fig:structure_evolution_overview}.
The details concerning the physics governing this evolution are developed in the next section.

% \onecolumngrid

\begin{table*}
\centering
\renewcommand{\arraystretch}{1.5}
\begin{tabular}{l|cclccc|cc|ccc|}
\cline{2-12}
                                    & \multicolumn{6}{c|}{\textbf{Control}}                                                                                                                                                                                                                                                                                                                                    & \multicolumn{2}{c|}{\textbf{Growth rate}}                                                                    & \multicolumn{3}{c|}{\textbf{Simulation}}                                                                                   \\
                                    & \multicolumn{6}{c|}{\textbf{parameters}}                                                                                                                                                                                                                                                                                                                                 & \multicolumn{2}{c|}{\textbf{at $k_x = 5k_{\rm min}$}}                                                            & \multicolumn{3}{c|}{\textbf{mesh}}                                                                                         \\ \cline{2-12} 
                                    & \multicolumn{1}{c|}{$\bm{g}$}                                                         & \multicolumn{2}{c|}{$\bm{\sigma_{\rm n,n}} | \bm{\sigma_{\rm \phi,\phi}}$}       & \multicolumn{1}{c|}{$\bm{\sigma_{\rm n,\phi}}$}                            & \multicolumn{1}{c|}{$\bm{\sigma_{\rm \phi,n}}$}                                             & $\bm{D} | \bm{\nu}$                & \multicolumn{1}{c|}{$\underset{k_y}{\bm{\max}} \bm{\gamma}$} & $\underset{k_y}{\mathbf{argmax}} \bm{\gamma}$ & \multicolumn{1}{c|}{$\bm{L_x} | \bm{L_y}$}   & \multicolumn{1}{c|}{$\bm{N_x} | \bm{N_y}$} & $\bm{\Delta t}$                \\
                                    & \multicolumn{1}{c|}{$\left[\frac{\Omega_0}{\rho_0} \frac{e \phi_0}{T_e n_0} \right]$} & \multicolumn{2}{c|}{$\left[\Omega_0 | \frac{\Omega_0}{\rho_0^2}\right]$} & \multicolumn{1}{c|}{$\left[\Omega_0 \frac{n_0 T_e}{e \phi_0} \right]$} & \multicolumn{1}{c|}{$\left[\frac{\Omega_0}{\rho_0^2} \frac{e \phi_0}{T_e n_0} \right]$} & $\left[ \Omega_0 \rho_0^2 \right]$ & \multicolumn{1}{c|}{$\left[ 10^{-4} \Omega_0 \right]$}       & $[k_{\rm min}]$                                   & \multicolumn{1}{c|}{$\left[ \rho_0 \right]$} & \multicolumn{1}{c|}{$[-]$}                 & $\left[ \Omega_0^{-1} \right]$ \\ \hline
\multicolumn{1}{|c|}{\textbf{I}}    & \multicolumn{1}{c|}{$1.5 \times 10^{-4}$}                                             & \multicolumn{2}{c|}{$6 \times 10^{-5}$}                                  & \multicolumn{1}{c|}{0}                                                 & \multicolumn{1}{c|}{0}                                                                  & $10^{-2}$                          & \multicolumn{1}{c|}{$\approx 4.3$}                           & $\approx 9$                                   & \multicolumn{1}{c|}{\multirow{3}{*}{$256$}}  & \multicolumn{1}{c|}{\multirow{3}{*}{1024}} & \multirow{3}{*}{$0.25$}        \\ \cline{1-9}
\multicolumn{1}{|c|}{\textbf{DW}}   & \multicolumn{1}{c|}{0}                                                                & \multicolumn{2}{c|}{$6 \times 10^{-5}$}                                  & \multicolumn{1}{c|}{$6 \times 10^{-5}$}                                & \multicolumn{1}{c|}{$6 \times 10^{-5}$}                                                 & $10^{-2}$                          & \multicolumn{1}{c|}{$\approx 4.3$}                           & $\approx 6$                                   & \multicolumn{1}{c|}{}                        & \multicolumn{1}{c|}{}                      &                                \\ \cline{1-9}
\multicolumn{1}{|c|}{\textbf{DW+I}} & \multicolumn{1}{c|}{$1.5 \times 10^{-4}$}                                             & \multicolumn{2}{c|}{$6 \times 10^{-5}$}                                  & \multicolumn{1}{c|}{$6 \times 10^{-5}$}                                & \multicolumn{1}{c|}{$6 \times 10^{-5}$}                                                 & $10^{-2}$                          & \multicolumn{1}{c|}{$\approx 8.0$}                           & $\approx 7.5$                                 & \multicolumn{1}{c|}{}                        & \multicolumn{1}{c|}{}                      &                                \\ \hline
\end{tabular}
\caption{\label{table:drift_wave_interchange} 
Main information regarding the \textsc{tokam2d} simulation cases considered in this paper: interchange only ``I", drift-waves only ``DW" and combined interchange/drift-waves ``DW+I".
It includes the control parameters of the governing equations system Eqs(\ref{system: TOKAM2D}), the maximum growth rate (see \cite{Ghendrih2022} for details) where $k_{\rm min} \rho_0 = \pi/256$ and the spatiotemporal mesh.}
\end{table*}
% \twocolumngrid

% !TeX spellcheck = en_GB
%
%

%!!!!!!!!!!!!!!!!!!!!!!!!!!!!!!!! SECTION !!!!!!!!!!!!!!!!!!!!!!!!!!!!!!!!!!!!!!!!!!!!!!!!!!!!!!!!!!!!!!!!!!!!!!!!!!!!!!!!!!!!!!!!!!
\section{Relaxation of localized overdense structures}
\label{section:ballistic_density_propagation}

\subsection{Comparison under drift-waves and interchange instability at similar growth rate}

The comparison between interchange and drift wave over-density propagation is performed with the values of the control parameters given in Table \ref{table:drift_wave_interchange} such that the linear growth rates of the two instabilities are identical. 

\begin{figure*}
	\centering
    \includegraphics[width= 1.\linewidth]{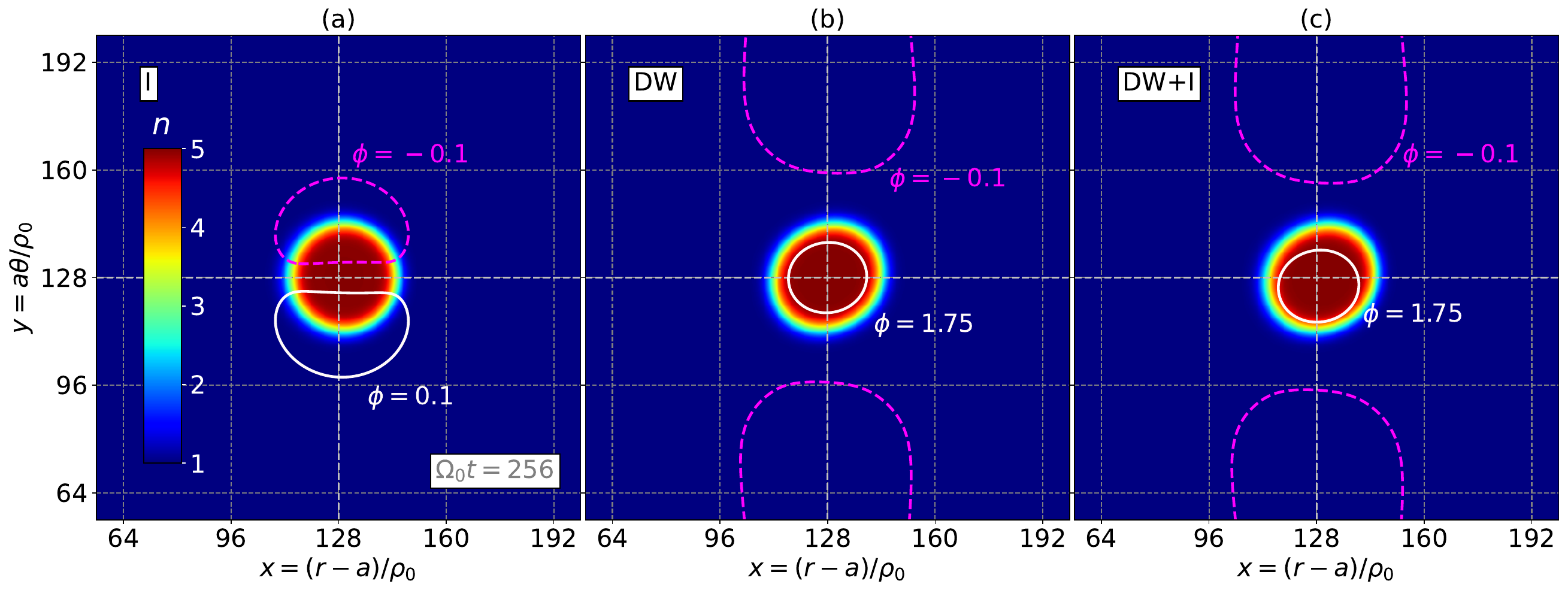}
	\caption{\label{fig:init_structure} \small Density at early time $\Omega_0 t = 256$ for each \textsc{tokam2d} case: interchange only (``I"), drift-waves only (``DW") and both instabilities combined (``DW+I").
    The initial density conditions are given in Table \ref{table:simulation_initial_conditions_blob} and the sets of control parameters in Table \ref{table:drift_wave_interchange}.}
\end{figure*}
The Fig.\ref{fig:init_structure} shows the density structure at an early stage, $\Omega_0 t = 256$, for each case: interchange only (``I"), drift-waves only (``DW") and both instabilities combined (``DW+I").
The density evolution governed by the interchange, Fig.\ref{fig:init_structure}\textcolor{blue}{a}, looks rather similar to the drift-waves case, Fig.\ref{fig:init_structure}\textcolor{blue}{b}, despite a slight outward shift in the former case and a weak poloidal narrowing in the latter case.
The main difference is in the potential structure.
For the drift-wave case, the component $\phi=n$ is large, so the negative values of the electric potential are distributed in the poloidal direction and are constrained by charge conservation. 
For the interchange case, a poloidally symmetric potential dipole with respect to the $x_{\rm od}$ axis is centered near the initial overdense region at $y_{\rm od}$.
The case combining both instabilities, displayed in Fig.\ref{fig:init_structure}\textcolor{blue}{c}, appears as a superposition of these two effects, such that the overdense region is slightly shifted radially while the potential is slightly shifted poloidally with respect to the drift-waves case.
Knowing that particle trajectories follow iso-potential contour, one can readily predict the next state of the density evolution.
With \textsc{tokam2d} convention, particles travel along negative - resp. positive - iso-potential in the counter-clockwise - resp. clockwise - direction in the $(x,y)$ plane.
In the interchange case, the potential structure in this early stage then acts as a rolling machine that pushes the density in the outward radial direction.
In the drift-wave case, one can consider two dipoles: the one in the upper half-plane above $y=y_{\rm od}$ and the one beneath it.
In the upper plane, the density is pushed inward while, in the lower plane, it's pushed inward.
Such a configuration then leads to a clockwise rotation of the overall overdense region.
The combined case is a bit more tricky to predict, due to the observed asymmetry between the potential and the density.
The situation on the potential is similar to the drift-wave case with a negative poloidal shift, while the overdense structure is still located poloidally around $y=y_{\rm od}$.
This imbalance favors a greater part of the density to undergo the dipole action for the dipole located at the highest poloidal coordinate.
These qualitative predictions only hold for a short amount of time, as the potential evolves with the density.

\begin{figure*}
	\centering
    \includegraphics[width= 1.\linewidth]{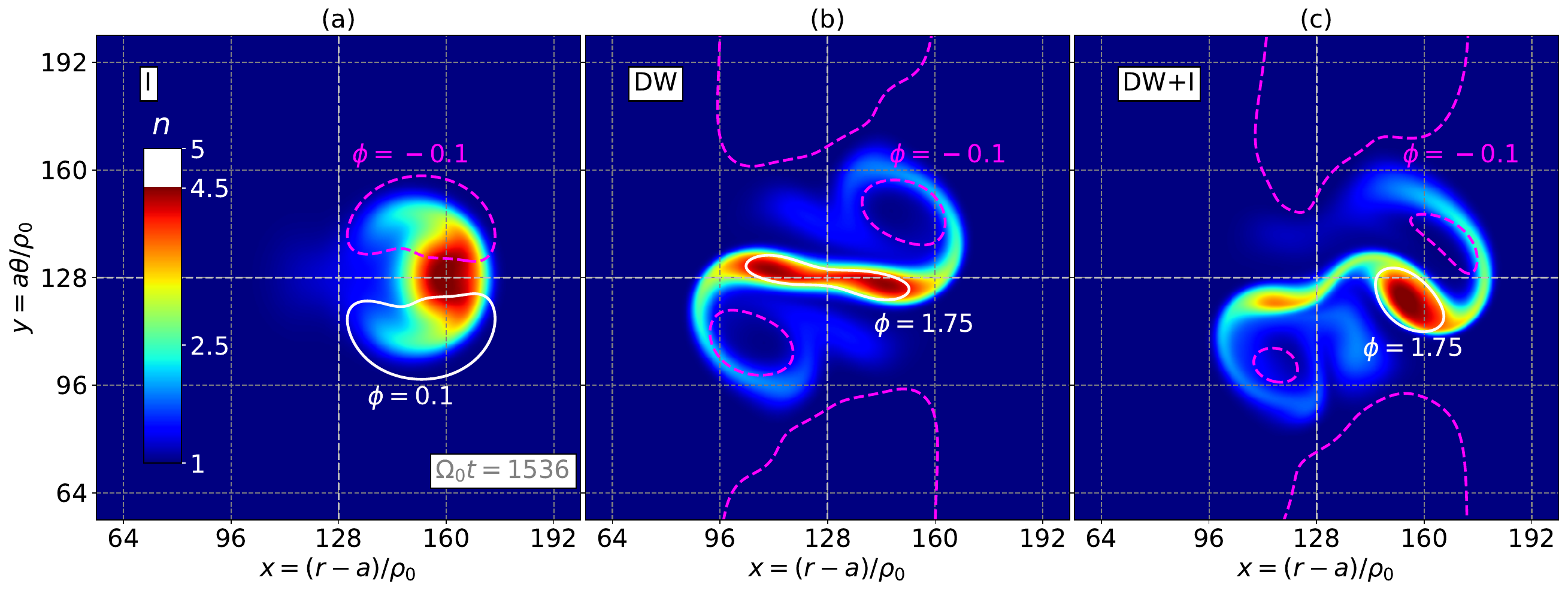}
	\caption{\label{fig:interm_structure} \small Density at time $\Omega_0 t = 1536$ for each \textsc{tokam2d} case: interchange only (``I"), drift-waves only (``DW") and both instabilities combined (``DW+I").
    The initial density conditions are given in Table \ref{table:simulation_initial_conditions_blob} and the sets of control parameters in Table \ref{table:drift_wave_interchange}.}
\end{figure*}
At a later time $\Omega_0 t = 1536$, the difference in the behavior of the interchange and drift wave relaxation processes is quite apparent as evidenced in Fig.\ref{fig:interm_structure}. 
The interchange relaxation, displayed in the snapshot Fig.\ref{fig:interm_structure}\textcolor{blue}{a}, leads to radial transport - observed to be ballistic, as detailed further -  with a displacement of the overdense region that preserves the initial axial symmetry along the $y=y_{\rm od}$ axis, as also observed in \cite{Yu2003} consistently with the electric potential dipole that follows the bulk of density. 
Conversely, the drift wave relaxation, snapshotted in Fig.\ref{fig:interm_structure}\textcolor{blue}{b}, does not initiate any global radial displacement in a preferred direction but rather exhibits a swirling pattern with a point symmetry with respect to the center of the initial overdense region $(x_{\rm od}, y_{\rm od})$.
This spiraling density ``brings" along the potential, changing the configuration such that the initial overdense regions begin to split into two distinct entities.
Both of these structures are consistent with the qualitative description made from the early stage.
Accordingly, a superposition of behavior of the two previous cases is observed in the case combining both instabilities, depicted in Fig.\ref{fig:interm_structure}\textcolor{blue}{c}.
Indeed, as discussed in the early stage, a shift between potential and density is responsible for a force due to dipoles being unevenly spread on the overdense region.
As for the drift-waves case, the spiraling causes the initial overdense region to split into two structures.
The difference in this case is that one of these structures carries a bigger part of the density.
In the meantime, the buoyancy driven by the interchange drive introduces an outward drift towards increasing values of $x$.
A satisfying remark at this point is that the additive ``basic" effects of the translation due to interchange and rotation due to drift-wave appear a great tool to explain the seemingly complex structure in the combined case in Fig.\ref{fig:interm_structure}\textcolor{blue}{c}.
Another interesting observation is that the maximum density has decreased between the early phase at $\Omega_0 t = 256$ and this intermediate phase at $\Omega_0 t = 1536$, while the mean density in the whole simulation domain is exactly the same in each simulation at all time.
This confirms that both interchange and drift-waves instabilities tend to restore thermodynamic equilibrium.

\begin{figure*}
	\centering
    \includegraphics[width= 1.\linewidth]{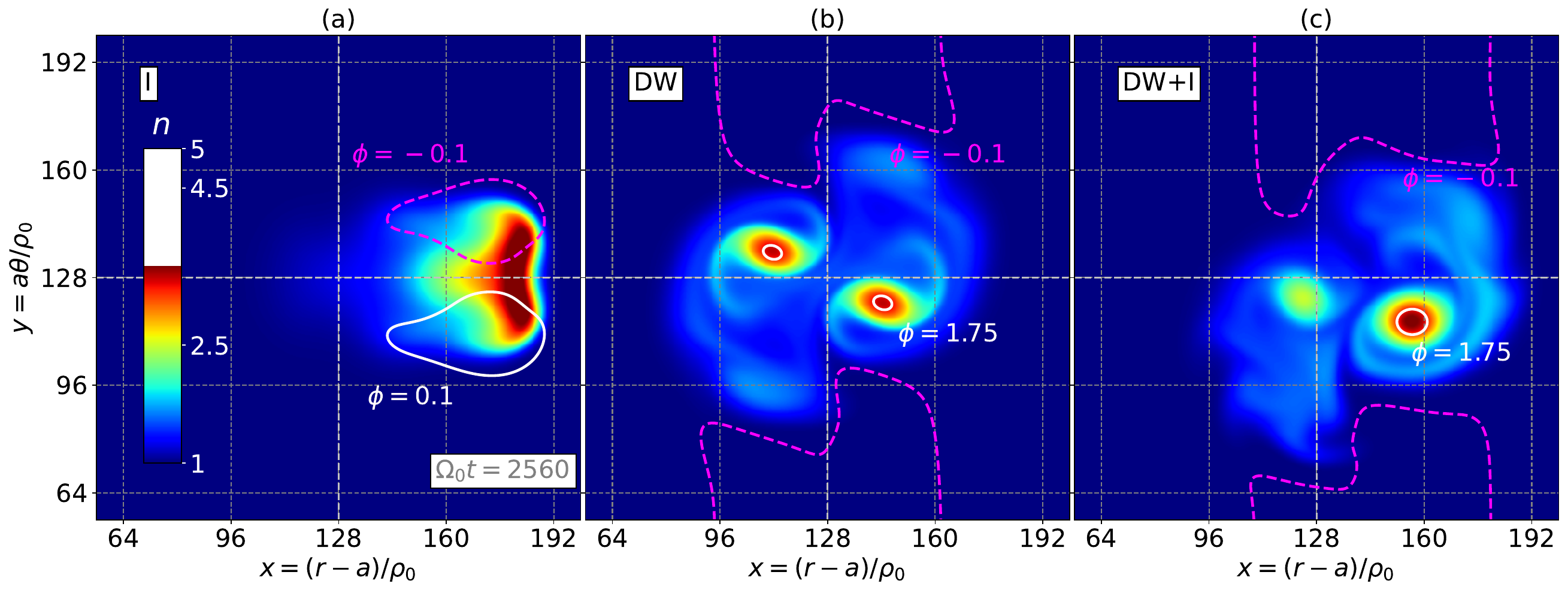}
	\caption{\label{fig:interm_structure2} \small Density at time $\Omega_0 t = 2560$ for each \textsc{tokam2d} case: interchange only (``I"), drift-waves only (``DW") and both instabilities combined (``DW+I").
    The initial density conditions are given in Table \ref{table:simulation_initial_conditions_blob} and the sets of control parameters in Table \ref{table:drift_wave_interchange}.}
\end{figure*}

At an even later time $\Omega_0 t = 2560$, the two aforementioned structures in the simulations including the drift-waves instability are clearly distinguishable while the case with interchange begins to exhibit a structure splitting itself, as shown in Fig.\ref{fig:interm_structure2}.
For the interchange case, shown in Fig.\ref{fig:interm_structure2}\textcolor{blue}{a}, the overdense region above - resp. below - the $y=y_{\rm od}$ poloidal coordinate exhibits a coherent shift in the increasing - resp. decreasing - poloidal direction. 
The reason is that, as the particles follow the iso-potential contour, the density is continuously distributed poloidally on both sides of the $y=y_{\rm od}$ line.
In the early stages, this motion manifested mainly as a radial translation of the initial overdense structure.
However, as the particles spread along the dipole, two coherent structures emerge.
In the two simulations including the drift-waves instability, displayed in Fig.\ref{fig:interm_structure2}\textcolor{blue}{b} and Fig.\ref{fig:interm_structure2}\textcolor{blue}{c}, the aforementioned splitting of the overdense region is finished and two structures are clearly distinguishable.
Regarding the drift-wave case in Fig.\ref{fig:interm_structure2}\textcolor{blue}{b}, on either side of the $x=x_{\rm od}$ line, the situation is somewhat analogous to the early stage as displayed in Fig.\ref{fig:init_structure} where each structure is associated to a potential leading to a rotation.
This suggests that, if the density gradient associated with each of these structures is strong enough, the rotating/splitting process could repeat for each of these substructures.

\begin{figure*}
	\centering
    \includegraphics[width= 1.\linewidth]{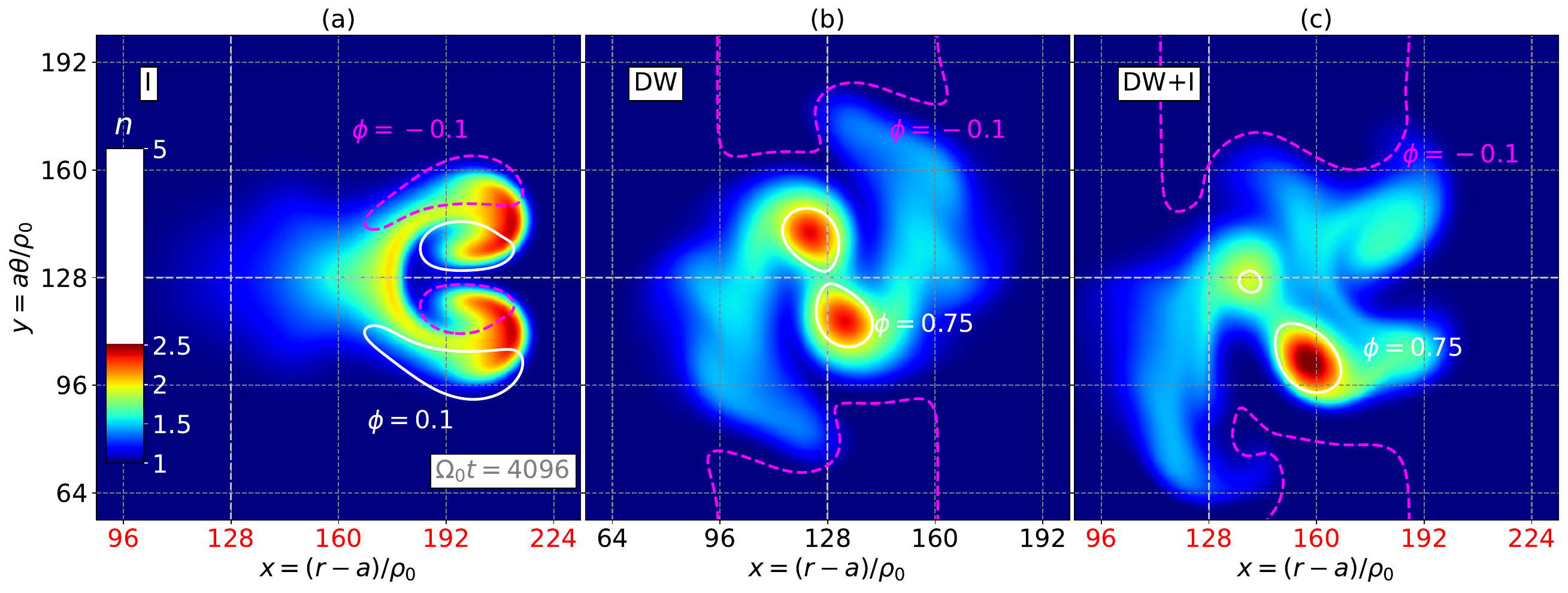}
	\caption{\label{fig:end_structure} \small Density at end time $\Omega_0 t = 4096$ for each \textsc{tokam2d} case: interchange only (``I"), drift-waves only (``DW") and both instabilities combined (``DW+I").
    The initial density conditions are given in Table \ref{table:simulation_initial_conditions_blob} and the sets of control parameters in Table \ref{table:drift_wave_interchange}.}
\end{figure*}
At the final time $\Omega_0 t = 4096$, the maximum density decreased by half-compared to the initial stage for each case, as shown in Fig.\ref{fig:end_structure}.
After this phase, there is little change in the structure's evolution, and the governing equations slowly restore a density of 1 throughout the simulation domain.
In the interchange case, displayed in Fig.\ref{fig:end_structure}\textcolor{blue}{a}, the splitting of the initial overdense region into two substructures poloidally distributed is achieved.
Each of these substructures is associated with its own potential dipole, such that each of them pursues its radial shifts.
While the drift-wave case - displayed in Fig.\ref{fig:end_structure}\textcolor{blue}{b} - did not evolve much compared to the previous snapshot in \ref{fig:interm_structure2}\textcolor{blue}{b}, an interesting feature is that the two substructures seem to converge towards one another.
This is explained by the initial rotating dynamic being slowed down while the density continues spreading. 
As the positive potential lies close to the density, this density spreading scatters the overdense regions and thus brings the iso-potential of each structure together.  
For the case combining both instabilities, the radial drift due to the interchange drive becomes less effective as observable in Fig.\ref{fig:end_structure}\textcolor{blue}{c}, where the density center of mass spanned a distance inferior to the interchange-only case as displayed in Fig.\ref{fig:end_structure}\textcolor{blue}{a}.
In that case, the coupling terms between density and potential then act as a drag for the transport induced by interchange.
In addition, this case is - among the two others - the one with the coherent structure carrying the most particles.
Indeed, as mentioned previously, combining both interchange and drift-waves instabilities leads to a poloidal symmetry breaking between density and potential, allowing for an imbalance of particle content in the substructures.

\begin{figure*}
	\centering
    \includegraphics[width= 1.\linewidth]{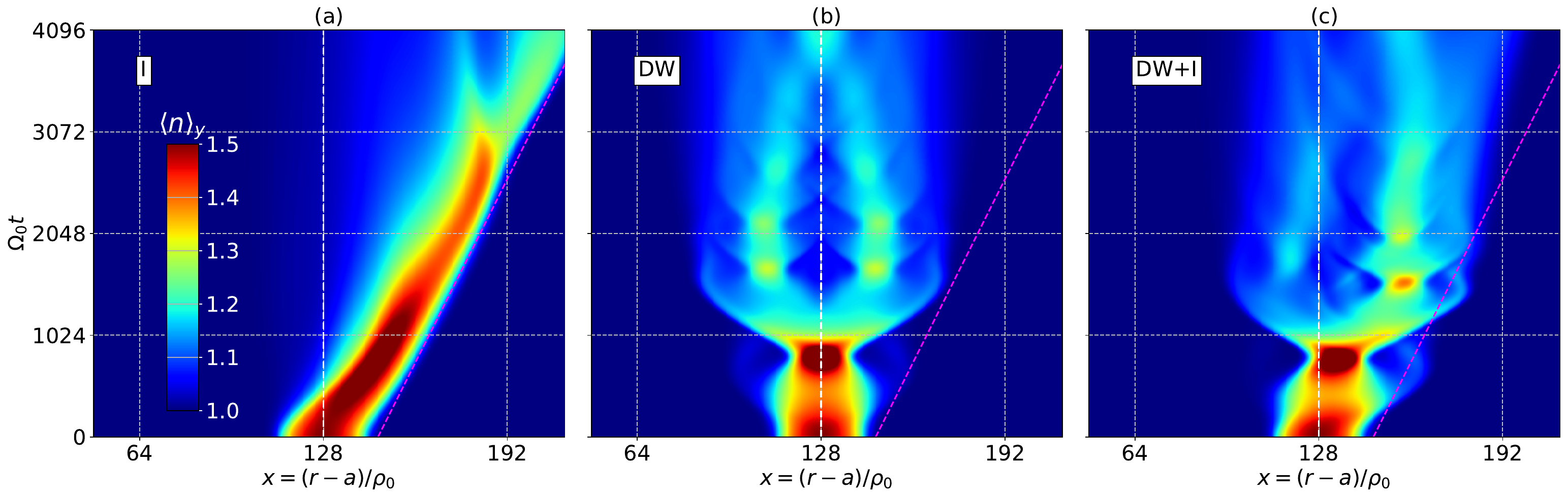}
	\caption{\label{fig:rtmap_nFSavg} \small Spatiotemporal evolution of the poloidally averaged density $\langle n \rangle_y$ for each \textsc{tokam2d} case: interchange only (``I"), drift-waves only (``DW") and both instabilities combined (``DW+I").
    The initial density conditions are given in Table \ref{table:simulation_initial_conditions_blob} and the sets of control parameters in Table \ref{table:drift_wave_interchange}.}
\end{figure*}

Another perspective on the previous observations can be gained by examining the spatiotemporal evolution of the poloidally averaged density $\langle n \rangle_y$, as illustrated in Fig. \ref{fig:rtmap_nFSavg}.
For the interchange relaxation displayed in Fig.\ref{fig:rtmap_nFSavg}\textcolor{blue}{a}, the motion of the initial the density forefront is ballistic motion, with a noticeable event at time $\Omega_0 t\approx 2800$ when the overdense region splits into two substructures.
The evolution of the density forefront can be fitted by the pink dotted line in Fig.\ref{fig:rtmap_nFSavg}\textcolor{blue}{a}, which corresponds to a velocity of $0.02 c_0$.

The behavior is altogether different for the drift wave relaxation displayed in Fig.\ref{fig:rtmap_nFSavg}\textcolor{blue}{b}, where the density remains symmetric with respect to the vertical axis at $x=x_{\rm od}$, therefore without displacement of the overdense mass center.
The previous observations, i.e. the splitting into two structures during the rotation and then their merging are clearly visible.
For the case combining both instabilities, depicted in Fig.\ref{fig:rtmap_nFSavg}\textcolor{blue}{c}, the appearing pattern is closer to the drift-waves case than the interchange case, with a clear imbalance for the outward substructure.
Compared with the interchange case, the radial motion of the overdense region is slowed down and ends faster.
This detailed analysis of the structure evolution under the influence of the drift-waves and/or interchange instabilities is useful for developing an intuition on the turbulent motion of coherent structures.
With these observations in mind, the contributions of the overdense region to the radial flux and zonal flow for each case are now analyzed.

\begin{figure*}
	\centering
    \includegraphics[width= 1.\linewidth]{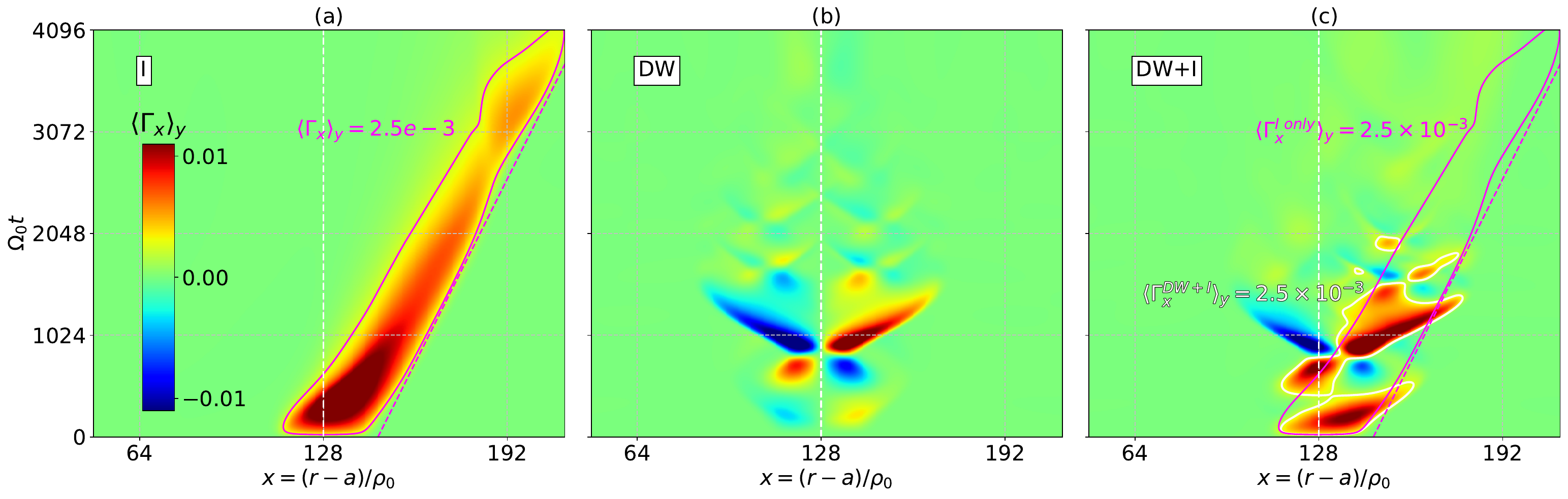}
	\caption{\label{fig:rtmap_flux} \small Spatiotemporal evolution of the poloidally averaged radial flux $\langle \Gamma_x \rangle_y$ for each \textsc{tokam2d} case: interchange only (``I"), drift-waves only (``DW") and both instabilities combined (``DW+I").
    The initial density conditions are given in Table \ref{table:simulation_initial_conditions_blob} and the sets of control parameters in Table \ref{table:drift_wave_interchange}.}
\end{figure*}

The spatiotemporal evolution of the radial particle flux averaged in the poloidal direction $\left\langle \Gamma_x\right\rangle_y$, where $\Gamma_x = - n \partial_y \phi$, for each case is shown in Fig.\ref{fig:rtmap_flux}. 
In the interchange case, displayed in Fig.\ref{fig:rtmap_nFSavg}\textcolor{blue}{a}, a radial outward drift of the transport pattern following the ballistic density displacement (highlighted with the pink dot line) is observed.
Conversely, for the drift-waves case displayed in Fig.\ref{fig:rtmap_nFSavg}\textcolor{blue}{b}, the spiraling motion yields a transport pattern with alternating particle flux from radially outward (red contours) to inward (blue contours), with no net radial shift of the pattern. 
Combining the two relaxation mechanisms leads to an alternating transport pattern biased to the outward motion and transport as shown in Fig.\ref{fig:rtmap_flux}\textcolor{blue}{c}.
On this same figure, the pink iso-contour is for the interchange case at $\left\langle \Gamma_x\right\rangle_y = 2.5 \times 10^{-3}$ to be compared to the white contour line at the same value but with the two relaxation mechanisms at play. 
The ballistic trend is comparable, but short-lived and partly balanced by inward particle flux. \\

\begin{figure*}
	\centering
    \includegraphics[width= 1.\linewidth]{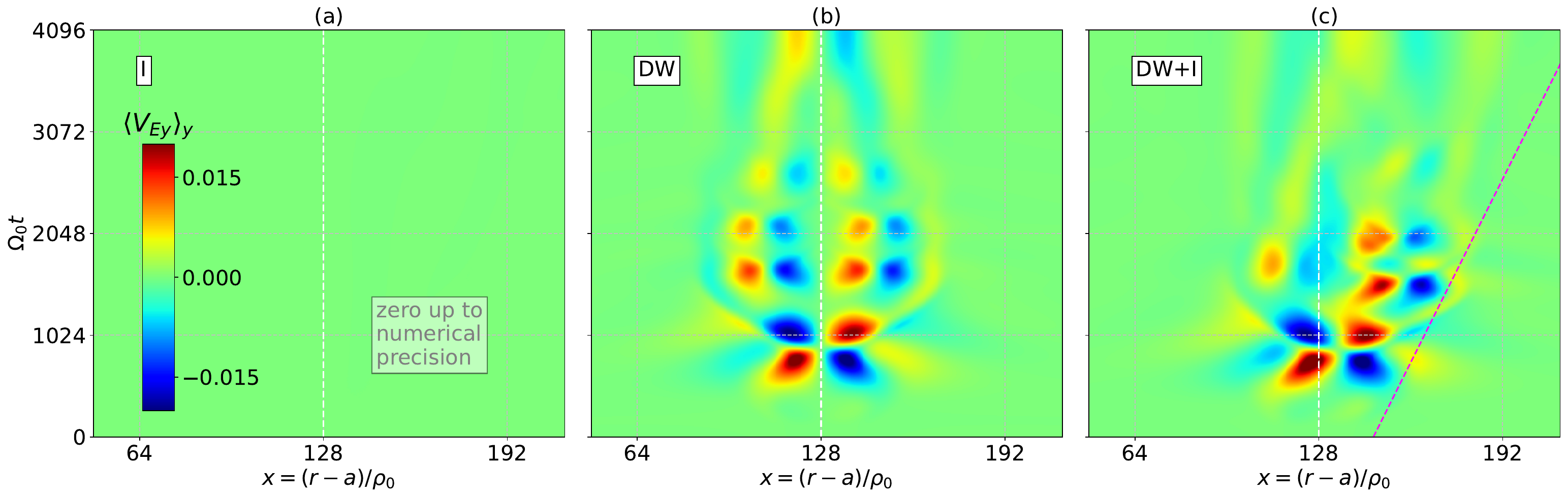}
	\caption{\label{fig:rtmap_VEy} \small Spatiotemporal evolution of the poloidally averaged poloidal electric drift component $\left\langle V_{Ey} \right\rangle_y$ - i.e. zonal flows velocity - for each \textsc{tokam2d} case: interchange only (``I"), drift-waves only (``DW") and both instabilities combined (``DW+I").
    The initial density conditions are given in Table \ref{table:simulation_initial_conditions_blob} and the sets of control parameters in Table \ref{table:drift_wave_interchange}.}
\end{figure*}

The contribution of each structure to zonal flows for each case can also be studied.
Their generation is observed through the poloidal averaged of the poloidal component of the electric drift $\langle 
V_{Ey} \rangle_y$, displayed in Fig.\ref{fig:rtmap_VEy}. 
For the interchange case, Fig.\ref{fig:rtmap_VEy}\textcolor{blue}{a}, the zonal flow generation is rigorously zero as the initial poloidal symmetry of the potential is conserved.
% \textcolor{red}{(Il doit y avoir un moyen simple d'expliquer ça avec les équations mais je n'ai pas encore trouvé.)}
Note that the interchange instability alone can generate zonal flow as long as a poloidal asymmetry on the density or potential - even of low amplitude - is present.
For the drift-waves case, Fig.\ref{fig:rtmap_VEy}\textcolor{blue}{b}, one can observe a quasi-symmetric pattern with an oscillatory behavior.
The maximum magnitude of the zonal flow is found to be comparable to the ballistic motion described above, typically $\sim 0.02 c_0$. 
The combined case is similar to the drift-waves only case, with the exception that the zonal flows undergo a slight radial shift because the potential lies close to the density coherent structure. 
Interestingly, the most dense substructure is associated with a stronger zonal flow generation.

\subsection{Interchange dominated case}

\begin{figure*}
	\centering
    \centerline{
        \includegraphics[width=1.0\linewidth]{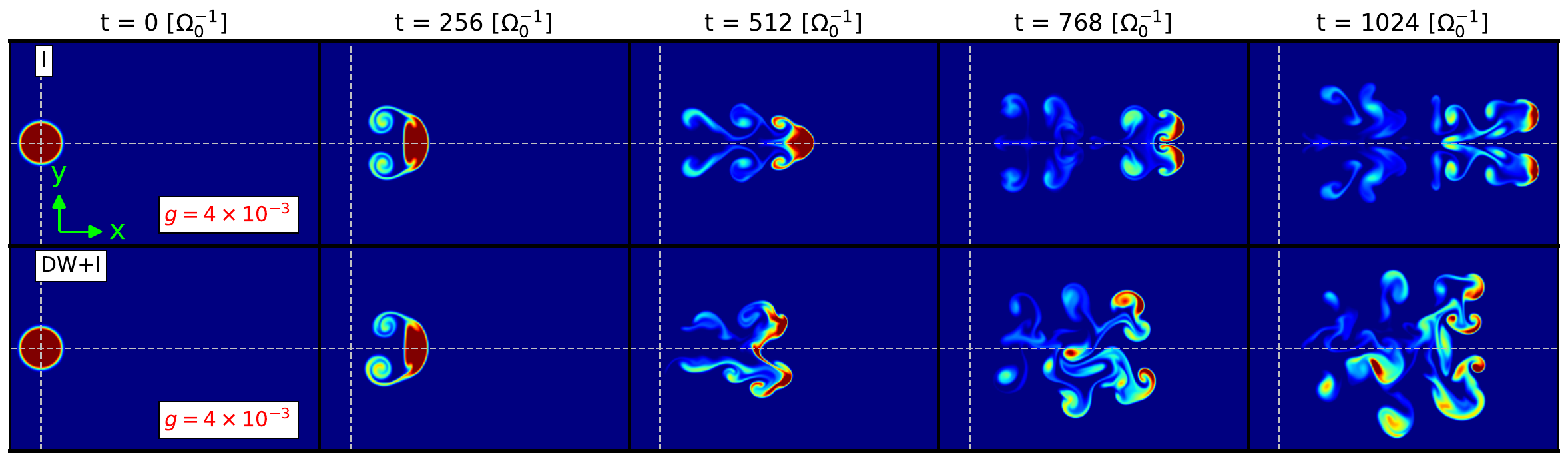}
    }
	\caption{\label{fig:structure_evolution_overview_interchange_dominated} \small Overview of the overdensity's structure evolution obtained with \textsc{tokam2d} with the parameters defined in Table \ref{table:drift_wave_interchange} with the exception of the interchange parameter set to $g=4\times10^{-3}$.
    ``I" stands for interchange only and ``DW+I" is a combination of interchange and drift-waves.}
\end{figure*}

While the previous cases - where each instability yields the same growth rate - are a great framework to study the effect of drift-waves and interchange in a controlled way, it appears that the drift-waves instability dominates the overall behavior of the structure relaxation.
One consequence is that the case combining both instabilities appears quite different from the typical ``mushroom" shape of blobs observed in experiments, e.g. see \cite{Maqueda2010, Katz2008}, and which is generally qualitatively retrieved in similar simplified models \cite{Easy2014, Angus2012, Angus2012a}.
This suggests that the contribution of the interchange to the growth rate should be greater than the one due to drift-waves to be closer to SOL conditions.
In consequence, to complete this study, a new case - analogous to the combined drift-waves/interchange ``DW+I" case (see Table \ref{table:drift_wave_interchange}) - is run with the interchange parameter increased to $g=4\times10^{-3}$ and with the overdensity initialized at $x_{\rm od}=32 \rho_0$.
For easier interpretation, this same case without density/potential coupling, i.e. without drift-waves instability, is also considered.
The density structure evolution for these two new cases is displayed in Fig.\ref{fig:structure_evolution_overview_interchange_dominated}.

\begin{figure*}
	\centering
    \centerline{
        \includegraphics[width=1.0\linewidth]{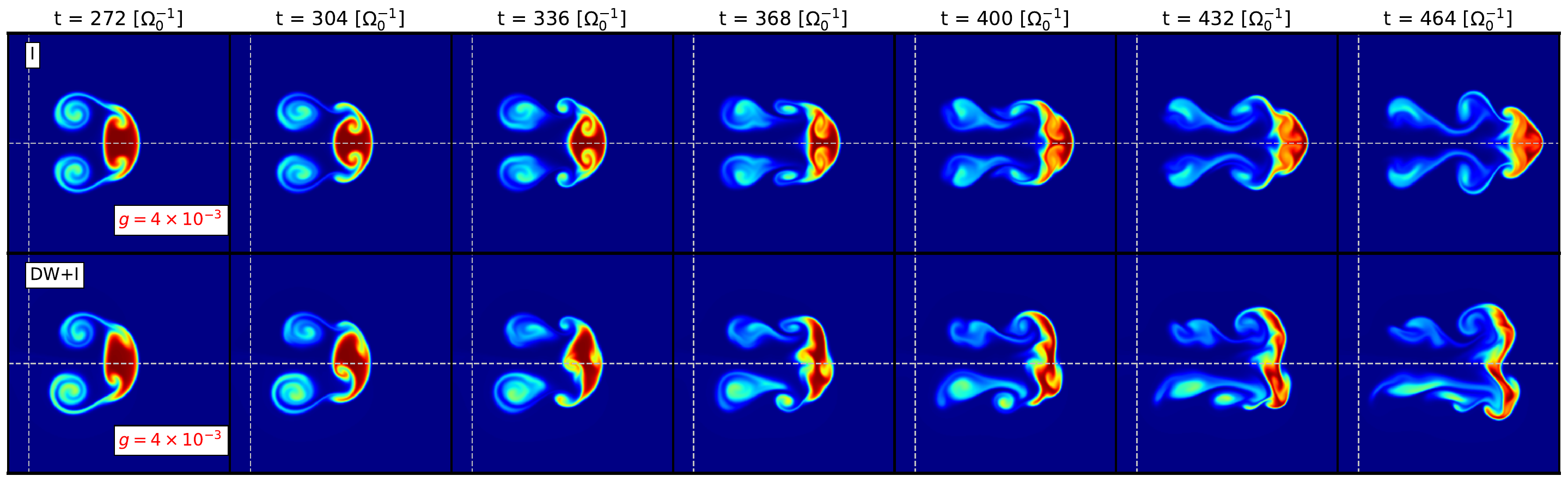}
    }
	\caption{\label{fig:structure_evolution_overview_interchange_dominated_splitting} \small Overdensity's structure evolution during the splitting of the structure of the ``DW+I" case obtained with \textsc{tokam2d} with the parameters defined in Table \ref{table:drift_wave_interchange} with the exception of the interchange parameter set to $g=4\times10^{-3}$.
    ``I" stands for interchange only and ``DW+I" is a combination of interchange and drift-waves.}
\end{figure*}

As expected, the dynamics - in particular the radial transport due to the interchange - is faster.
For both cases, in the early stage of the evolution, one can recognize the typical ``mushroom" shape characteristic of these SOL transport models \cite{Angus2012, Easy2014}.
Here the boosted interchange parameter accounts for spiraling wings to develop along the radial path of the overdense structure.
This feature was not observable in previous cases, likely because of the particle spreading along the iso-potential being damped by the density diffusion term in Eq(\ref{equation: evolution n}).
This behavior appears discrete in time, with chunks of density detaching from the bulk structure at a somewhat regular frequency.
Apart from this feature, the case with only interchange is quite similar to the one studied in the previous section, with a conserved poloidal symmetry and the initial overdense region that gradually splits into two substructures.
The case with both instabilities, but dominated by interchange, appears drastically different from the ``DW+I" case studied earlier.
In the early stage, as expected, only a slight poloidal asymmetry in the density is noticeable with, this time, the radial drift overcoming the previously dominating rotation of the overdense region.
However, at an intermediate stage $\Omega_0 t \approx 400$, the density bulk splits into two similar substructures with irregular shapes.
This splitting, which occurs early compared to the case with only interchange, can be qualitatively appreciated in Fig.\ref{fig:structure_evolution_overview_interchange_dominated_splitting} which displays the structure evolution during this event.
This closeup look indicates that the splitting is preceded by a rotating motion of the density bulk, characteristic of the drift-wave instability, accelerating the initial structure splitting.
A remarkable feature is that, compared to the case combining both instabilities with similar growth rates, the resulting substructures are about symmetric poloidally with respect to the $y=y_{\rm od}$ line. 

\begin{figure*}
	\centering
    \includegraphics[width= 0.75\linewidth]{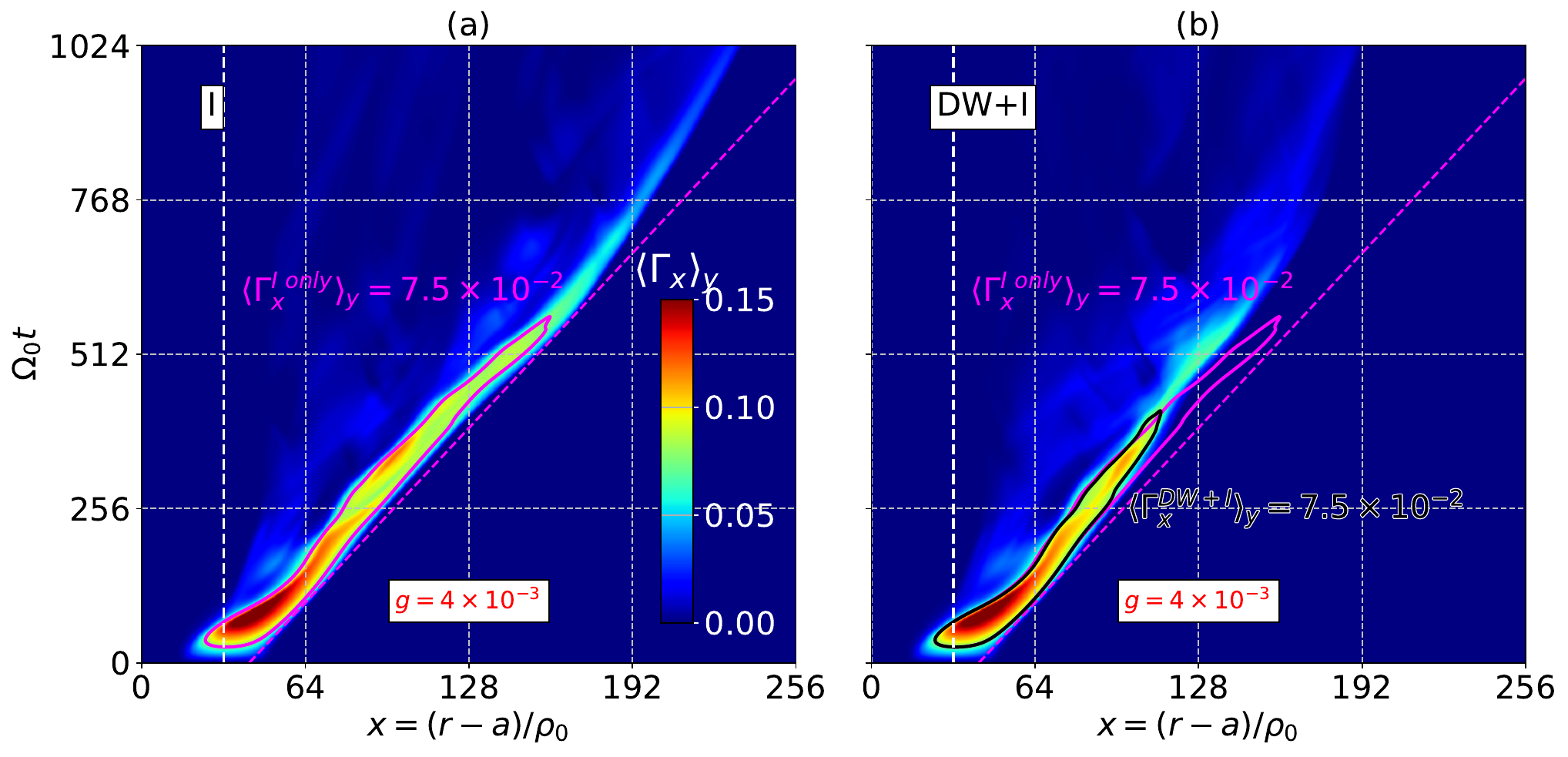}
	\caption{\label{fig:rtmap_flux_FSavg_interchange_dominated} \small Spatiotemporal evolution of the poloidally averaged radial flux $\langle \Gamma_x \rangle_y$ for each \textsc{tokam2d} case: interchange only (``I"), drift-waves only (``DW") and both instabilities combined (``DW+I").
    The initial density conditions are given in Table \ref{table:simulation_initial_conditions_blob} and the sets of control parameters in Table \ref{table:drift_wave_interchange}.}
\end{figure*}

Concerning the mean flux, displayed in Fig.\ref{fig:rtmap_flux_FSavg_interchange_dominated}, the characteristics observed in the previous sections hold.
In the case with only interchange, shown in Fig.\ref{fig:rtmap_flux_FSavg_interchange_dominated}\textcolor{blue}{a}, the transport is outward and ballistic, as observed experimentally \cite{Choi2024}.
The associated velocity associated with the pink dotted line representing the ballistic motion is about $\sim 0.2 c_0$, i.e. about 10 times the one observed previously.
The spiraling wings detaching from the overdense structure are also noticeable in this plot and carry a small portion of the overall transport.
For the case combining both instabilities, displayed in Fig.\ref{fig:rtmap_flux_FSavg_interchange_dominated}\textcolor{blue}{b}, the transport appears similar to the case with only interchange in the early stage, but progressively slows down and ends earlier.
This behavior is consistent with the structure evolution in Fig.\ref{fig:structure_evolution_overview_interchange_dominated}, the drift-waves instability seems to effectively break big structures into smaller ones while spreading them poloidally.

\begin{figure*}
	\centering
    \includegraphics[width= 0.5\linewidth]{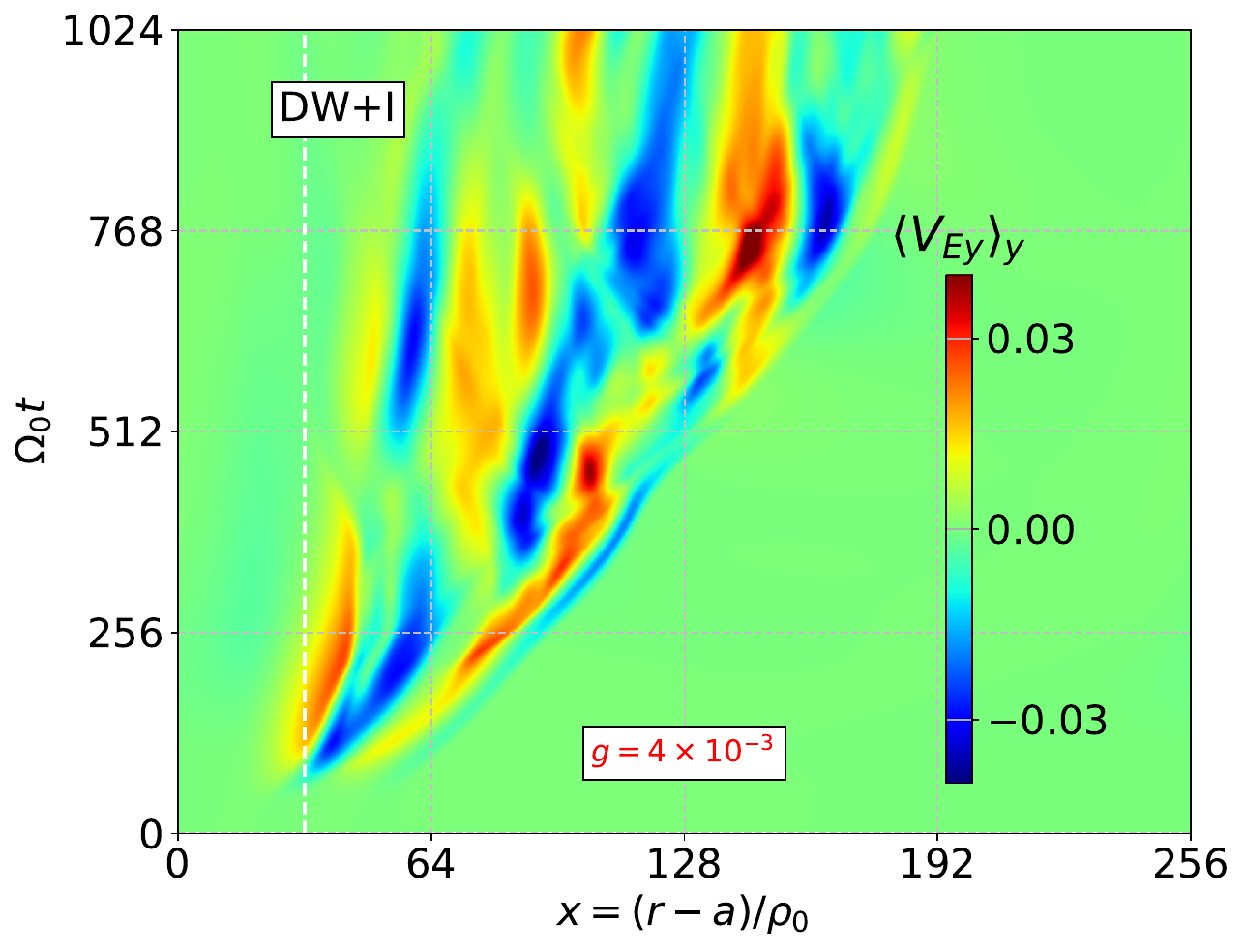}
	\caption{\label{fig:rtmap_VEy_interchange_dominated} \small Spatiotemporal evolution of the poloidally averaged poloidal electric drift component $\left\langle V_{Ey} \right\rangle_y$ - i.e. zonal flows velocity - for each \textsc{tokam2d} case: interchange only (``I"), drift-waves only (``DW") and both instabilities combined (``DW+I").
    The initial density conditions are given in Table \ref{table:simulation_initial_conditions_blob} and the sets of control parameters in Table \ref{table:drift_wave_interchange}.}
\end{figure*}

Finally, the zonal flow evolution of the case combining both instabilities can be observed in Fig.\ref{fig:rtmap_VEy_interchange_dominated}.
Note that the case with only interchange, as in the previous case with a weaker interchange parameter, does not generate any zonal flow.
While the previous case combining both instabilities with a weaker interchange drive exhibited a short-lived oscillatory behavior of zonal flow generation (see Fig.\ref{fig:rtmap_VEy}\textcolor{blue}{c}) this new case is different.
Indeed, here the zonal flow generation is uninterrupted, due to the chunk of density detaching from the bulk of density that keeps propagating outward.
As the drift-waves instability breaks the poloidal symmetry of the initial overdense region, the interchange instability contributes to the generation of finite zonal flows.

\section{Discussion and conclusion}
\label{section:discussion_and_conclusion}
Single relaxation events have been analyzed under drift-waves and interchange instability. 
This investigation is based on numerical simulations using the code \textsc{tokam2d} and evolution equations of the density and electric potential in a 2D domain with conditions akin to SOL. 
Simplifying the parallel loss terms, in particular by retaining only the linearised form, allows one to monitor the instability mechanism driving turbulent transport, either considering a single drive or combining them. 
The linear analysis of the equations yields the dispersion relation and the instability growth rate.
This analysis is used to determine the sets of control parameters to be used in the simulations. 
Parameters are chosen to enforce identical growth rates for the interchange and drift-waves instabilities \cite{Ghendrih2022}. 
It is to be underlined that forms of the Kelvin-Helmholtz instability can also play a role in the evolution of the $E\times B$ flows.\\

The present model couples the density field to the electric potential with equations that stem from the particle and charge conservation equations, in that, following the pioneer papers \cite{Garbet1991, Nedospasov1993, Hasegawa1978}. 
If wished, one can readily substitute the density with the electron thermal energy and recover similar equations. 

The single relaxation events are obtained with prepared initial conditions, a circular high-density region in the plane transverse to the magnetic field. 
Without an external source, this overdense region relaxes towards the background uniform density. 
With the interchange instability only, the electric drift governs a displacement of the overdense structure together with a change of its shape with the main density peak narrowing in the radial direction and extending poloidally. 
The electric potential dipole that onsets the $E \times B$ motion is a signature of the mechanism at work. 
At a later stage, the relaxation pattern exhibits a structure of the electric potential in the wake of the motion, and a splitting of the main density peak at the forefront. 
The initial radial motion of the density forefront is ballistic.
In the absence of an initial poloidal density imbalance, this relaxation event generates zero net zonal flows.

Under the influence of drift-wave instability alone, the overdense region exhibits a distinct behavior: it rotates around the positive peak of the electric potential, which was aligned with the density peak.
The spinning overdense region generates spiraling arms of density, spreading out the density radially and poloidally, but without enforcing a net displacement of the mass center. 
Eventually, the overdense structure breaks into two coherent substructures, rotating around the same initial peak of positive potential.
An alternating short-lived zonal flow pattern is generated.

When both instabilities are accounted for, with each contributing equally to the total growth rate, a mix of the relaxation processes associated with isolated instabilities is observed.
A slight initial misalignment of the electric potential on the overdense region generates a response that is typical of the drift wave relaxation process, but with a poloidal of density for each arm - en eventually each substructure - that develops during the spinning motion.
On top of that, a net radial displacement of the density peak is observed, which is reminiscent of the interchange relaxation process. 
Compared with a situation with only the interchange instability, this motion is slower and put to an end way more quickly.
Consequently, this combined case generates less transport than the case with only interchange even though the maximum growth rate is twice as big.
The zonal flow pattern is similar to the one due to the drift-waves instability, and slightly favors positive amplitudes. \\

Finally, a situation considering both instabilities but where the interchange dominates is considered to be more representative of experimental observation, where a typical ``mushroom" shape of blobs is witnessed.
Two new main features are observed.
The first feature is the intermittent deposition of chunks of density along the rapid radial motion of the overdense region through the spiraling motion of ``wings" developing at each poloidal end of the bulk structure.
This behavior is due to the boosted interchange drive allowing a faster redistribution of particles along the electric potential dipole.
The second feature is the splitting of the initial overdense regions into two substructures, resulting from the spinning due to the drift-waves instability, which almost preserves a poloidal symmetry.

This work reveals intricate interactions between drift-wave and interchange instabilities, which can mostly be explained by the behavior of each instability taken separately.
In addition, this work is also meant as a pioneering study for establishing a systematic numerical method for detecting structures in turbulent plasmas.
Structure identification is appealing for filtering turbulent events in simulations and experimental data. 

This research is supported by the National Research Foundation, Singapore.
\bibliographystyle{apsrev4-2}
% \bibliography{bibliography,Article_Blob}
\bibliography{Article_Blob}

\appendix

\end{document}